# RESEARCHES of ALPHA and BETA RADIOACTIVITY at LONG-TERM OBSERVATIONS

A.G.Parkhomov

Institute for Time Problem Researches. The Lomonosov Moscow State University, Moscow, Russia.

http: // www.chronos.msu.ru

The part of results obtained on multichannel installation created for researches of various processes, including α and β decays, in a combination to recording of environment parameters, is represented. The installation works more than 13 years practically continuously. At the count rate measurements of the $^{60}Co$ and $^{90}Sr$-$^{90}Y$ β sources the rhythmic changes with amplitude 0.3 % from average magnitude and period 1 year, and up to 0.02 % with period about one month are detected. Count rate rhythmic changes of α sources $^{239}Pu$ are not detected. The results obtained by us are compared to similar results of other researchers. The argumentation of critics is analyzed.

Until recently was considered, that the nuclear decays are caused by exclusively *intranuclear* processes, on which course usual external influences (electromagnetic, thermal, acoustic, etc.) cannot noticeably influence. Therefore, at measurements of a radioactivity, it should be observed and it was really observed only exponential drop with chaotic fluctuations corresponding to the Poisson distribution. But recently, when it has become possible to spend long-term exact measurements, the experimental results indicating presence of rhythmic and sporadic deviations from an exponential drop are obtained. The hint on a possibility of deviations from the exponential law of a radioactive decay was obtained, for example, at long measurements having by the purpose the definition of half-life of long leaving radionuclides [1]. The results obtained on our installation, specially created for realization of long-term measurements, confirm presence of such effects.

**Experimental installation**

In itself, detection of *variations of radioactivity measurements* does not mean a proof of *variations of a radioactivity*. The observable deviations are very small and slow. Therefore, it is necessary for tracing on stretch of many days, months, years. At such precision measurements the large statistics and minimization of instability of registering equipment is required.

In order to results of measurements could enough confidently be interpreted just as variations of a radioactivity, the experimental installation permitting long time measuring of a counting rate from several α and β sources is created. The exterior factors which can influence results (temperature, atmospheric pressure, humidity, radiation background, voltage of power system) are inspected. The detectors, rather stabiles and possessing practically unlimited resource - halogen G-M counters were applied to registering a β and γ radiation, the silicon detectors were applied to registering an α of particles. The detectors with radioactive sources and power supplies are located in the thermostat. The installation works practically continuously on stretch 13 years. More detailed definition of experimental installation and techniques of registration of signals are given in papers [2, 3, 4].

**Basic results**

Figure 1 shows, as varies the count rates from $^{60}Co$ and $^{90}Sr$-$^{90}Y$ β sources, measured by various detectors on stretch more than 10 years. Rhythmical variations with amplitudes of 0.3 % from the average value with the period of a 1 year are obvious. The comparison of average courses of a count rate and temperature near the installation (Fig.2) shows different dynamics of year cycles. Other basic parameters of an exterior medium – radiation background, atmospheric pressure and humidity of an air, power supplies differently behave also. It allows to state, that detected variability does not grow out influences of the usual factors of an environment.

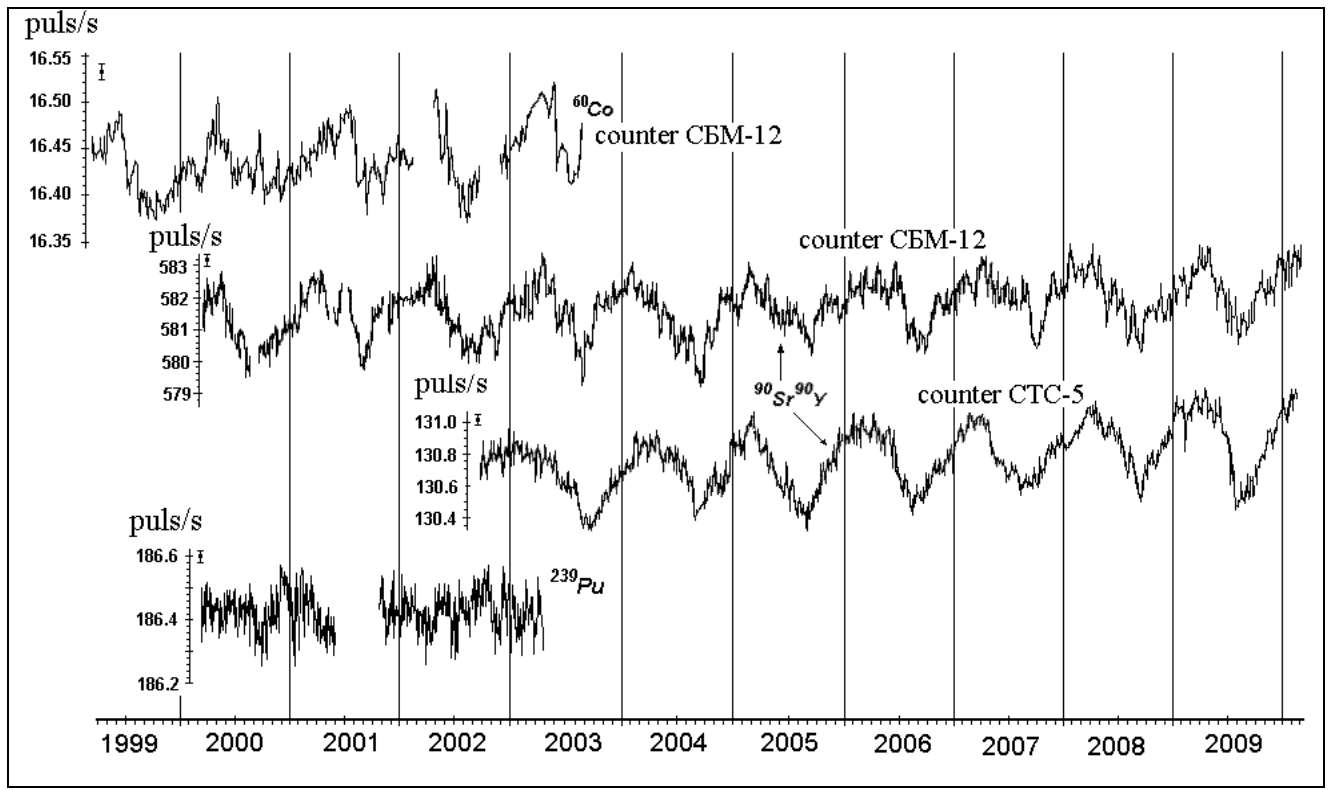

**Fig. 1.** Count rate of the $^{60}Co$ and $^{90}Sr$-$^{90}Y$ β sources, measured by G-M counters, adjusted for a drop of activity with half-lifes 5,27 and 28,6 years, and count rate of the $^{239}Pu$ α source, measured by the silicon detector [3, 5].

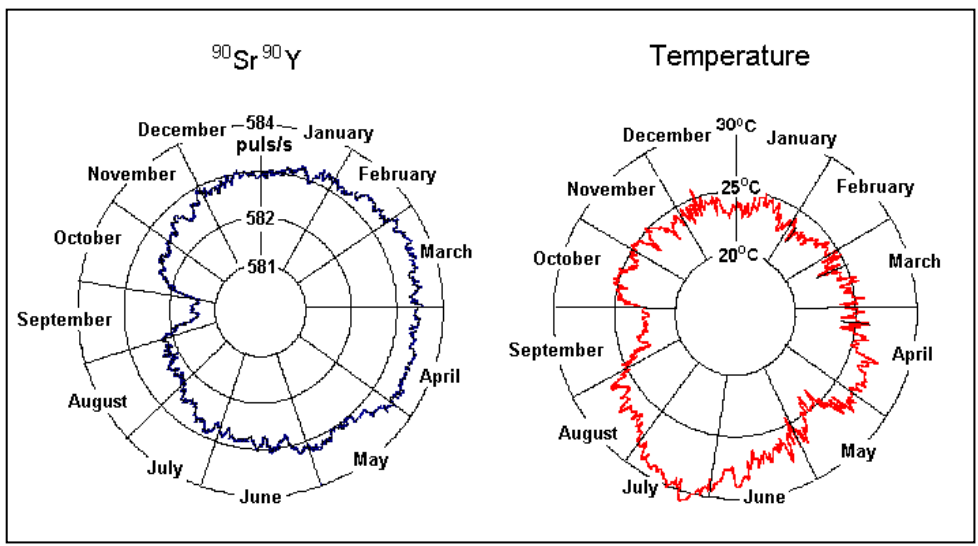

**Fig. 2.** Comparison of an average count rate of $^{90}Sr$-$^{90}Y$ β source to a course of temperature about installation. The results obtained since April 2000 till March 2007 are treated.

Large duration of measurements and sufficient statistics allow to apply the frequent analysis and to represent periodograms. On the periodogram of the count rate of devise with $^{90}Sr$-$^{90}Y$ source (Fig.3) the 1-year period and it harmonics (182, 91.5, 61 days) is most seen. The period of lunar month (29.27 days) is shown. Especially clearly this rhythm is seen at an average of results on cycles of synodical lunar month. About new moons the count rate on the average on 0.02 % is more than about full moons (Fig. 4).

In area of near 1 day period, peak of sun day with a thin structure reflecting interaction of this rhythm with a year rhythm and its harmonics is seen clearly. The peak appropriate to lunar day (1.0375) is seen. The amplitude of near day variations does not exceed 0.003% from average value and, in difference from variations with 1-year and monthly periods, it is impossible with confidence to state, that they are not generated by temperature influences.

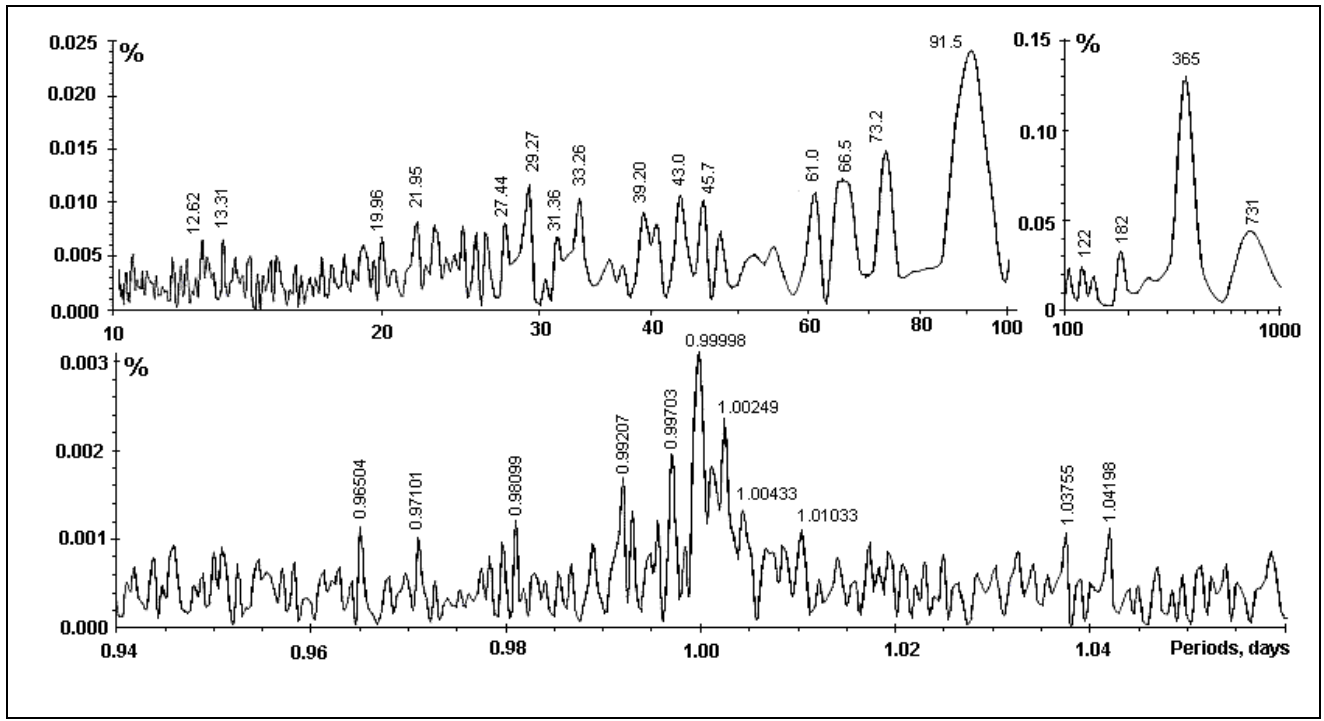

**Fig. 3.** Periodograms of count rate variations of the $^{90}Sr$-$^{90}Y$ β source with the G-M counter СБМ-12. Count rate analyzable time interval since April 2000 till March 2007. Amplitude - in percentage of an average count rate [3, 5]. The periods appropriate to tops are shown.

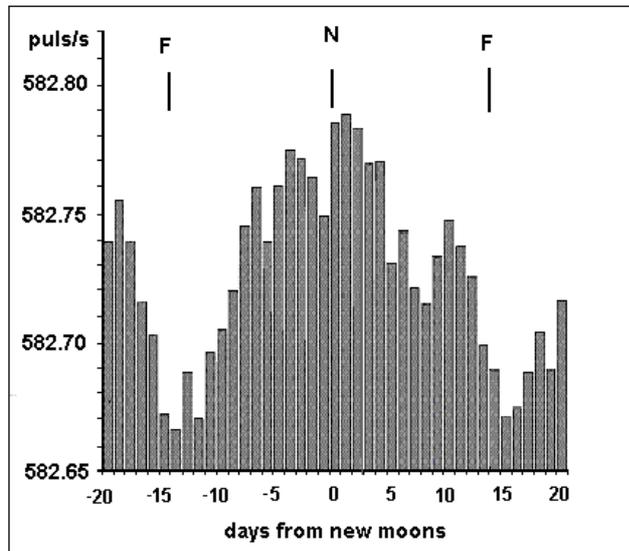

**Fig. 4.** 87 cycles of synodical lunar month average of the $^{90}Sr$-$^{90}Y$ β source count rate with the G-M counter СБМ-12 (April 2000 - March 2007). N – new moon, F – full moon.

It is important to note, that the rhythmic variations in these experiments are found out only at measurements of a β radioactivity. The similar researches of $^{239}Pu$ α radioactivity with use of the silicon detector, practically insensitive to a β and γ radiation, do not reveal rhythmic variations of count rate. The observable chaotic fluctuations with amplitude of the order 0.1% from an average count rate (see fig. 1), apparently, are connected to a noise of the silicon detector and recording electronics engineering. This conclusion is confirmed by results of long-term α particles registration of one $^{239}Pu$ source by two silicon detectors (Fig.5): everyone from detectors fluctuates in own way. In difference from it, synchronism of variations at a measurement of β radioactivity of different sources or of one source by two different type's detectors (see fig. 1 and 8) testifies that only in case the β radioactivity a change of a velocity of a radioactive decay takes place.

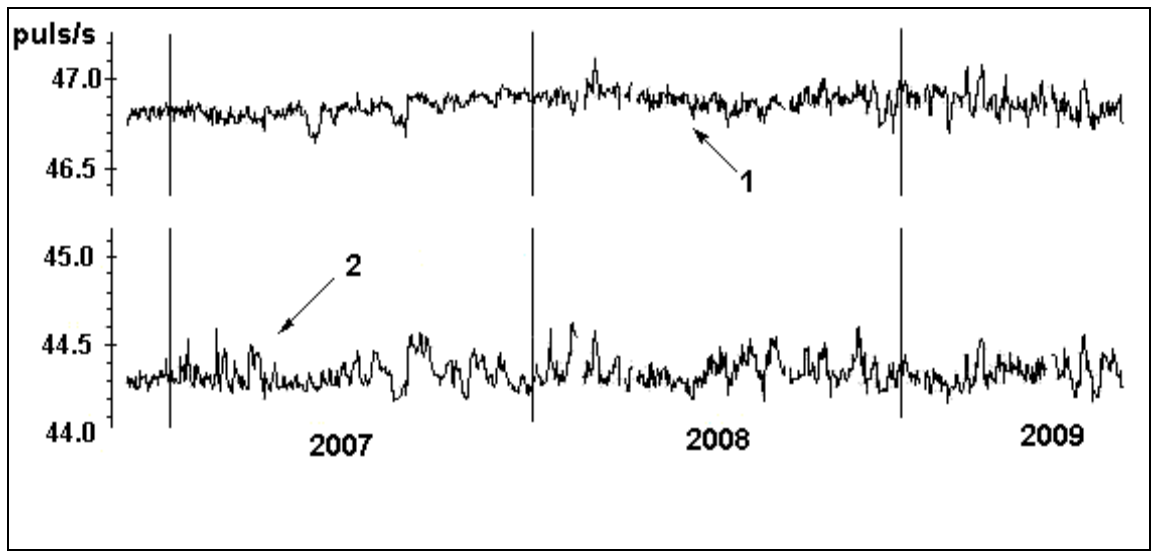

**Fig. 5**. Count rates of one $^{239}Pu$ α source, measured by two silicon detectors.

**Experiments of other researchers**

The annual rhythm is found out in Troitsk at spectrometric measurements of tritium β decay with the purpose of the electronic antineutrino mass definition [6] (Fig. 6). Month and diurnal rhythm is found out at measurements of gamma radiation $^{60}Co$ and $^{137}Cs$ by the NaJ(Tl) detector (Dubna) [7]. In paper [8] (Dubna and Troitsk) the presence of diurnal variations in results of $^{60}Co$ and $^{137}Cs$ γ radiation measurements by Ge(Li) detector are revealed. To these dates, however, it is necessary to concern with caution, as their magnitude (up to 0.7 %), is improbably great. At our measurements, the amplitude of diurnal changes of count rates does not exceed 0.003 % for $^{90}Sr$-$^{90}Y$ source and 0.01% for $^{60}Co$ source.

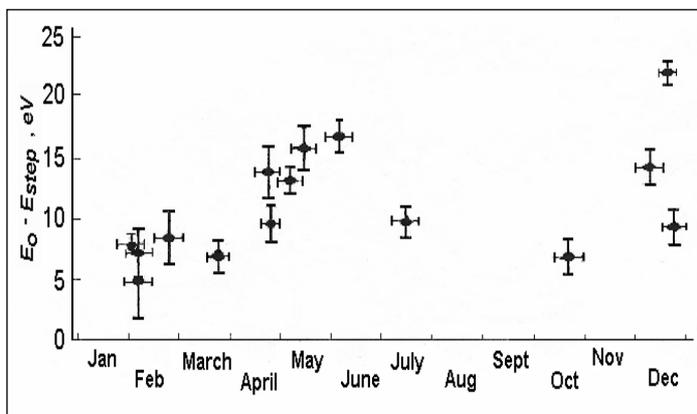

**Fig. 6.** Experiment in Troitsk [6]. Annual changing of difference between theoretical and experimental segments of a spectrum of tritium β decay near to maximum electron energy (1994-2001).

Annual rhythm of a radioactivity is exhibited at measurements of half-lifes long living radionuclides. This effect was exhibited clearly at the definition of $^{32}Si$ half-life [1] (Brookhaven National Laboratory). For measurements a β activity the gas-flow proportional counter was used. The results of measurements by duration more than four years are shown in a fig. 7. Simultaneous measurements on the same counter of activity slowly decays $^{36}Cl$ (a half-life $3·10^5$ years) had the purpose compensating of instability of the measuring equipment. It is visible, that the count rate of both radionuclides synchronously varies with amplitude 0.2 – 0.3 % and period 1 year.

The rhythmic deviations from average value of approximately same amplitude with a period 1 year are found out at 15-year's $^{226}Ra$ radioactivity measurements with use of the ionization chamber (Fig. 8) [11] (Physikalisch-Technische-Bundesandstalt in Germany).

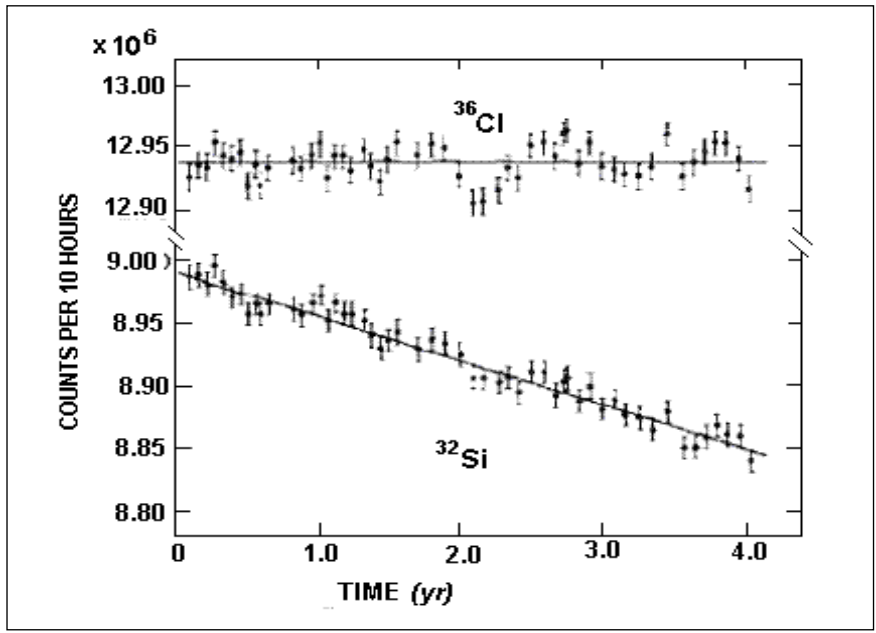

**Fig. 7.** Periodic changes of count rate of $^{32}Si$ and $^{36}Cl$ [1]

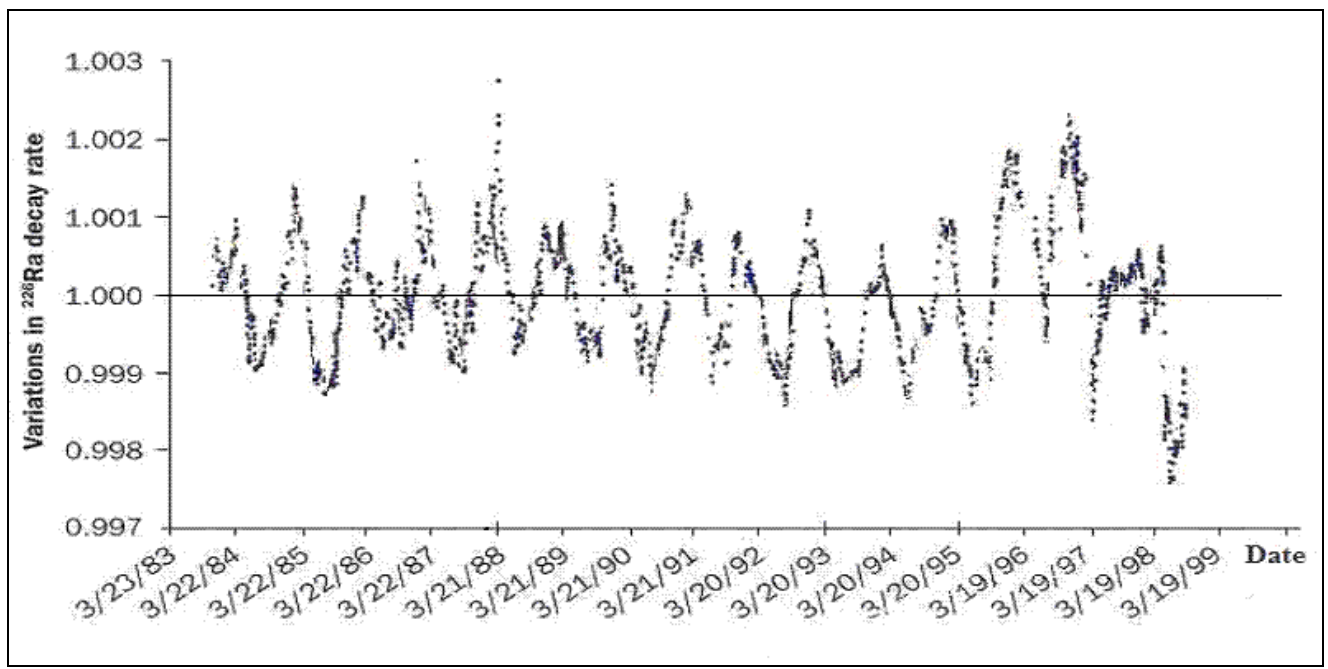

**Fig.8.** Deviations of $^{226}Ra$ radioactivity measurements from average value [11].

### Discussion

The doubts of such deviations from the conventional representations about a radioactivity are quite natural. Let's consider the argumentation of the opponents explained in papers [12, 13].

**1.** At measurements by germanium detector of ratios $^{22}Na/^{44}Ti$, $^{241}Am/^{121}Sn$ and $^{133}Ba/^{108}Ag$ γ radiation count rate authentic modifications with 1-year period is not revealed. Because of these dates is concluded a lack of effect of rhythmic variations of a radioactivity [12].

But the constancy of fraction not necessarily means an invariance of a numerator and denominator. It can be connected to uniformity modifications of a registered count rates. And the reason uniformly influencing to results of measurements, is not necessary connected to instability of the measuring equipment. Therefore lack of variations of a ratio of radioactivities does not mean a lack of variations of radioactivities of separate radionuclides. Instability of an equipment and action of varying temperature, pressure, humidity etc. are very various in different laboratories. Nevertheless, periods, phase and magnitude of effect at measurements of different radionuclides in various laboratories with use unequal equipment are very close. It indicates existence of the nontrivial reason uniformly influencing to activity of various radionuclides in a different time and in different places.

2. In paper [13] data from the of the $^{238}Pu$ radioisotope thermoelectric generators aboard the Cassini spacecraft are analyzed. At a variation of a distance up to the Sun from 0.7 up to 1.6 a.u. the difference from an exponential curve does not exceed 0.01%. On this foundation idea about connection changes a radioactivity with a distance between the Earth and Sun [9, 10] are concluded an inaccuracy.

But, as the power output $^{238}Pu$ practically completely is connected with α decays, the results of analysis of Cassini spacecraft power output means "correct" course only *α decays*, and are good acknowledgement of a conclusion on a lack of noticeable anomalies in a course α decays already made at our experiments [3, 5].

All dates, known to us, on anomalies of a radioactivity course are connected to a *β* radioactivity. On the first sight, this statement contradicts annual rhythmic, found out at a measurement of activity $^{226}Ra$ [11] (Fig 8). But $^{226}Ra$ is **not only** *α* source, as it generates a long chain containing not only *α*, but also β decays. For radioactivity measurement $^{226}Ra$ the ionization chamber was used. It is detector, sensing to β and γ radiations. Usually $^{226}Ra$ sources are in hermetic ampoules which are not passing α particles. In this case registered effect is connected to a β and γ radiation completely. Therefore, the presence in radioactivity measurements $^{226}Ra$ of variations with 1-year periodicity can be connected with a β radioactivity. For a conclusion about presence such variations in α decays the experiment [11] does not give of any foundations.

So, the results of analysis of Cassini spacecraft power output can not refute idea about possible connection of β radioactivity with a distance between the Sun and Earth, because in α decays the considered effect is not exhibited. Other matter, that idea about connection of changes of radioactivity with 1-year period with Sun neutrino fluence oscillations [9, 10] looks extremely doubtful because of exclusive weakness of interaction such neutrino with substance. On the other hand, presence of that effect in β radioactivity and the lack it in α radioactivity, indicates to participation a neutrino to this effect (neutrino - necessary participant of β processes, but does not accept participation in α decays). It is possible that considered effect is connected with a stream of "relict" neutrino [14, 15]. A hypothesis about a probable role of "relict" neutrino offered also for explanation of the not clear effects which have been found out at a measurement of neutrino mass [6]. The substantiation of these ideas requires the special consideration.

---